\documentstyle[12pt,aaspp4,flushrt]{article}

\begin{document}
\title{STELLAR POPULATIONS AND VARIABLE STARS IN THE
CORE OF THE GLOBULAR CLUSTER M5\footnote{Based on observations with the NASA/ESA
Hubble Space Telescope, obtained at the Space Telescope Science
Institute, which is operated by AURA, Inc., under NASA contract NAS5-26555
}}

\author{Laurent Drissen}
\affil{D\'epartement de Physique and Observatoire du Mont M\'egantic, Universit\'e Laval, 
Qu\'ebec, QC, G1K 7P4\\Electronic mail: ldrissen@phy.ulaval.ca}
\and
\author{Michael M. Shara}
\affil{Space Telescope Science Institute, 3700 San Martin Drive, Baltimore, Maryland 21218\\
Electronic mail: mshara@stsci.edu}

\def\et{{\it et \thinspace al.}\ }

\begin{abstract}
We report the discovery of a variable blue straggler in the core
of the globular cluster M5, based on a 12-hour long series of images
obtained with the Planetary Camera aboard the {\it Hubble Space Telescope}.
In addition, we present the light curves of 28 previously
unknown or poorly studied large-amplitude variable stars (all but one are
RR Lyrae) in the cluster core. A (V, U-I) color-magnitude diagram shows
24 blue stragglers within 2 core radii of the cluster center.
The blue straggler population is significantly more centrally concentrated than
the horizontal branch and red giant stars.
\end{abstract}

\newpage
\section{Introduction}

Globular cluster cores are being imaged intensively with the Hubble Space
Telescope, and continue to yield surprises. The realizations that
blue stragglers are commonplace, red giants are depleted, and that horizontal
branch populations may depend on core properties has spurred on observers
in the past five years. As part of our program to study the distributions
of blue stragglers and variable stars in globular cores, we have now imaged
NGC 5904 = M5.

The globular cluster M5 harbors one of the richest collections of RR Lyrae
stars in the Galaxy. It is also home to one of only two known dwarf novae
in Galactic globular clusters. Until now, however, not a single blue straggler
candidate has been identified in M5. Is this because none exist, or
because blue stragglers reside only in the (until now unobservable)
core of M5?

In section 2 we describe the observations and reductions; the CMD of the
core of M5 is presented in section 3. The variable stars we find are described
in section 4 while our results are summarized in section 5.

\section{Observations and reductions}

\subsection{Search for variable stars}

A series of twenty-two 600-second exposures of M5's central (r$\leq 50''$) 
region, spanning 11.5 hours, was obtained on March 21, 1993 with the 
{\it Planetary Camera} (PC) aboard the {\it Hubble Space Telescope} (HST)
before the repair mission. 
The F336W filter, similar to the Johnson U bandpass (Harris {\et al.} 1991), 
was used for the observations.
Standard calibration (bias and dark subtraction, flatfield
correction) was performed with STScI software within
IRAF/STSDAS.

The detection of variable stars was performed as follows. First, the 22 
individual images were combined with IRAF's {\it combine} task (with the 
{\it crreject} option) to create a deep image. This image is shown in Figure 1.
The individual images were remarkably well aligned with each other, so
no shift was necessary. DAOPHOT's Daofind algorithm (Stetson 1987) was 
then used to build a masterlist of 3144 stars visible in the deep image.

In order to eliminate the numerous cosmic rays, which are very difficult to 
remove on single frames, the 22 individual images were combined two by two 
(again with the {\it combine} task and the {\it crreject} option), with an 
overlap of 1 (image 1 + image 2, image 2 + image 3, ...).
Aperture photometry with an aperture radius of 4 pixels (0.17$''$), and a sky 
annulus with inner and outer radii of 10 and 15 pixels was then performed on 
the 21 resulting images, {\it using the masterlist of stars previously found}. 
The standard deviation from the mean in the 21 frames was then computed for
all stars; this resulted in the discovery (or re-discovery) of 29 variable
stars, which will be discussed in section 4. 

Note that since for any given star the conditions (PSF, flatfield, star
location, aperture, background) are virtually identical from frame to frame
(with the exception of faint stars near bright, large amplitude variables),
the relative photometry from frame to frame is expected to be much more
precise than the relative photometry among stars within the same frame
(mostly because of PSF variations which induce position-
dependent aperture corrections and overlapping PSFs in a crowded field).
While the relative error from one star to the next within a single frame
can be as large as 0.2 mag (if simple aperture photometry is used), even for
bright stars, the typical standard deviation from the mean of the 21 frames
for a given non-variable star is less than 0.05 mag at U=17.5, and reaches
0.1 mag at the cluster main sequence turnoff.

\subsection{The color-magnitude diagram}

In order to study the stellar populations in M5, a 70 second F555W
(equivalent of Johnson V filter; Harris {\it et al.} 1991) and a 70 second 
F785LP (comparable to I) PC images were retrieved from the HST archives.
Both images were obtained on December 27, 1991.
Fortunately, the center of these frames is only $\sim$ 8$''$ away from the
center of our images (the total field of view of the 4 PC CCDs is 70$''$), so 
most stars detected in our F336W images are also located in the archive frames;
the area common to all three sets of images is $\sim 1.1$ arcmin$^2$.
Stellar photometry was performed in the most simple way on these frames:
aperture photometry (with the same parameters as above) was performed on
the deep, combined F336W frame using the masterlist. After proper shift
and rotation of the coordinates, the same masterlist was used to obtain
aperture photometry of the stars in the F555W and F785LP images.
Although PSF-fitting techniques are known to improve the photometric accuracy 
(especially
at the faint end of the luminosity function and near bright stars; see 
Guhathakurta {\it et al.} 1992 for a detailed discussion), simple core
aperture photometry gives results  which are reliable enough
(typically $\sigma \sim 0.2$ mag above the turnoff) for our purpose.
A total of 2153 stars have been included in the color-magnitude diagram
discussed below.

Aperture corrections and photometric zero-points determined by Hunter \et 
(1992) were used to calibrate the magnitudes and colors of the stars.
Moreover, the absolute sensitivity of the CCDs is known to decrease slowly
with time following each decontamination (Ritchie \& MacKenty 1993); this
effect was taken into account. 

As a consistency check, the photometric calibration was compared with 
ground-based data.
The average F555W and F336W magnitudes of the RR Lyrae stars are 
$F555W_{RR} = 15.2$ and $F336W_{RR} = 15.55$,
while Buonanno et al (1981) find $V_{RR} = 15.13$ and Richer \& Fahlman (1987)
obtained $U_{HB} = 15.6$. The agreement is excellent.

\section{Stellar Populations}

The color-magnitude diagram for the 2153 stars located in the area
common to the F336W (U), F555W (V) and F785LP (I) frames is shown in Figure
2. In this plot, variable stars (see next section) are identified with
special symbols; seven of 29 variables found in the F336W frames are
located outside the F555W and F785LP field of view, and are therefore
not included in Figure 2.
Photometry is fairly complete down to $\sim$ 0.5 magnitude below the
turnoff. The large wavelength difference between the U and I filters compensates
for the relatively poor photometric accuracy, and allows us to clearly
define the stellar populations. 
Two characteristics of the CMD are worth noting: the significant blue straggler
population and the morphology of the horizontal branch.

\subsection{Blue Stragglers} 

The boundaries of the BS region of the CMD have often been arbitrarily
defined, vary from author to author, and also depend on the filters used.
As emphasized by Guhathakurta \et (1994), the (V, U-I) CMD is well suited
to define the blue straggler population.
As a working definition, we decided to include in the BS region of the
(V, U-I) CMD all stars
significantly brighter (0.4 mag) than the turnoff and bluer than the
average main sequence at the turnoff level;
twenty-four stars are included in this region of the CMD (surrounded by
dotted lines in Figure 2) and can be considered as blue stragglers.
Such a significant blue straggler population in the core of M5 is in striking 
contrast to the outer regions of the cluster, where none have been found
(Buonanno {\it et al.} 1981, for $2' \leq r_{obs} \leq 5.6'$; Richer \& Fahlman 1987, for $3' \leq r_{obs} \leq 25'$). Recent CCD observations of the
central regions of M5 (Brocato \et (1995); Sandquist \et (1996)) also reveal
some blue stragglers.
The two-dimensional distribution of red giants, horizontal branch stars and 
blue stragglers is shown in Figures 3a-c. It is obvious from these plots that 
the BS population is preferentially concentrated in the very core of the 
cluster.
This is quantified in Figure 4a, which shows the cumulative distribution
of the different stellar populations within the inner 50$''$, and in
Figure 4b, within one core radius ($24''$).
A simple K-S test showed that there is a 99.5\% probability that the
blue stragglers in the core of M5 are more centrally concentrated than
the horizontal branch stars, and a 97.5\% probability that they are
more centrally concentrated than the red giants.

This tendency for blue stragglers to be more centrally concentrated than
other cluster members has been noted for most globular clusters observed
to date (see Sarajedini (1993), Yanny \et 1994 and references therein),
and is consistent with the hypothesis that BS are more massive than
main sequence stars.

The luminosity-inferred lifetime of the shortest-lived, most luminous blue straggler in the
core of M5 is $\sim 5 \times 10^8$ yrs, assuming it lies on the hydrogen-burning
main sequence. This corresponds to the star
with V = 16.0, $L \sim 25 L_\odot$ and $M \sim 2.15 M_\odot$. The relaxation
time in the core of M5 is $T_{rc} = 2 \times 10^8$ yrs (Djorgovski 1993).
Hence, all BS in the core of M5 are expected to be dynamically relaxed.
Thus, the radial distribution of the BS is, indeed, a direct probe of their 
masses.

In order to compare the blue straggler population from cluster to cluster,
Bolte, Hesser \& Stetson (1993) have defined the {\it specific frequency},
$F_{BS}$, as the ratio of the number of blue stragglers to the total
number of stars brighter than 2 magnitudes below the horizontal branch
at the instability strip. In the case of the central region of M5 (area
in common to the U, V and I HST/PC frames), where $V_{HB} = 15.2$,

$$F_{BS} = {24 \over 434} = 0.055 \pm 0.015$$

This value is smaller than the specific frequencies found by Guhathakurta
\et (1994) for the core of M3 (0.09) or 47 Tuc (0.07). There is no known
correlation 
between $F_{BS}$ and any obvious cluster property.

\subsection{Horizontal Branch Morphology}

The second important feature to notice in the CMD is that the horizontal 
branch is clearly split into two sub-populations, with a significant gap 
(0.25 mag) in between, at $V \sim 15.5, U-I \sim -0.2$. This gap is also 
clearly present in the CCD data of Brocato \et (1995) and possibly of Reid 
(1996), but not in the high-quality data of Sandquist \et (1996). The large
wavelength coverage of the WFPC2 data probably increases the width of the
gap. Hesser (1988) has pointed out that
discontinuous horizontal branches with very blue tails are surprisingly
common, including M15, NGC 288, NGC 1851, and at least 10 other clusters.
He suggests that gaps may provide important constraints on mixing in earlier 
stages of stellar evolution. We see no evidence that the gap width
changes between the inner and outer parts of the core.

\section{Variable Stars}

Figure 5 shows the standard deviation as a function of the average U magnitude 
for all the stars in the F336W frames. The usual problem with outliers
in Figure 5 (due to mismatches in star lists) was almost completely avoided
because (a) the same master list for all frames (see section 2.1), and (b)
all images were perfectly aligned with each other. Visual inspection
of the few remaining outliers immediately showed them to be mismatches and they
were eliminated from our photometry. Stars with RMS magnitude of
variability higher than 2 sigma above the mean curve were considered
possible variables and were examined more closely. Most of them were faint
stars located close to bright, large-amplitude variables.
Twenty-eight stars stand out as bright, large-amplitude variables 
in Figure 5, as well as one fainter, low-amplitude variable star.
Seven of these variable stars had never been identified previously, and
two (HST-V20 and V21) were considered as a single star.
All but one of the large-amplitude variables have light curves typical of
RR Lyrae stars, whereas the low-amplitude variable is a blue straggler.
Information on the variables, including cross-identification with
previous papers, is presented in Table 1. We note that the cataclysmic
variable V101 is $36''$ North and $282''$ West of the center of M5 and
is therefore too distant to have been included in the field of view of the
Planetary Camera ($70'' \times 70''$).

\subsection{ A Variable Blue Straggler}

Of 24 blue stragglers identified in Figure 2, only one (HST-V28) shows
evidence for variability above the noise level. Although only HST-V28 showed
up above the 2-sigma variability level in Figure 5, we have examined
the light curve of every blue straggler and compared them with those of nearby
stars of similar magnitude in order to find possible variations
undetected by our technique; no new variable showed up. But because the noise
level is still relatively high between V=17 and V=18 (the 2$\sigma$ variability
level varies from 0.06 to 0.12, corresponding to amplitudes
$\sim $ 0.2 - 0.3 mag), we cannot exclude the possibility that some blue
stragglers are actually small amplitude variables. 

The color-magnitude diagram (Figure 2) shows that HST-V28 is the bluest,
and one of the brightest blue stragglers in M5 (with $M_V = + 2.2$).
Figure 6 shows its light curve, along with light curves of 3
nearby comparison stars of the same magnitude. HST-V28 is obviously variable,
with an amplitude $\Delta U \geq 0.15$ mag. Unfortunately, our data
do not allow us to determine the period with any precision (or even to 
determine if the variations are periodic). Phase dispersion minimization 
algorithms favor a periodicity of the order of $\sim$ 15 hours (in which
case this star could be an eclipsing binary), but cannot 
completely exclude periods shorter than $\sim$ 1.9 hours (typical for
pulsating blue stragglers in clusters; Mateo 1993). Better photometric
accuracy and temporal sampling are obviously necessary to determine the period
and nature of HST-V28.

Guhathakurta \et (1994) recently discovered 2 variables among the 28
blue stragglers in the core of M3. Both have photometric amplitudes
$\sim 0.4$ mag and periods $\sim$ 6-12 hours. Guhathakurta \et suggest
that both stars are dwarf cepheids.

\vskip 1.0truecm
\subsection{RR Lyrae Stars}

Among the 97 genuine variables listed in Sawyer-Hogg's (1973)
catalogue, 93 are RR Lyrae. Recently, Kalda {\it et al.} (1987)
and Kravtsov (1988, 1992) searched for new variables in the central,
crowded region of the cluster and found 30 more. So far, the light
curves of only two of these new variables have been published
(Kravtsov 1992). A recent review of the M5 RR Lyrae population has been
published by Reid (1996).

The U-band light curves of all RR Lyrae stars found in the HST images
are shown in Figure 7.
Tentative periods have been determined for the 9 type RRc stars with a
phase dispersion minimization algorithm, and the results are presented
in Table 1. The periods of RRab stars are too long to be determined from
our observations alone.

\vskip 1.0truecm
\subsection{HST-V1}

Eclipsing binary stars are common in the periphery of M5's core (Reid 1996,
Yan \& Reid 1996), but star HST-V1 is unique among the variables found in the 
core of M5. Although
our observations do not cover an entire cycle (P$\sim 0.55^d - 0.65^d$),
the light curve suggests that HST-V1 is a contact (or semi-detached)
binary. The secondary minimum is very asymmetric, somewhat reminiscent of 
the near-contact binary V361 Lyrae (Kaluzny 1990). 
The strong asymmetry of the light curve of V361 Lyr, more pronounced at shorter
wavelengths, is thought to be caused by the presence of a hot spot on
the secondary star resulting from the high mass transfer rate from the
primary star. 

Unfortunately, HST-V1 lies outside the boundaries 
of the F555W and F785LP frames, so its color cannot be determined.
But with an average magnitude of U=15.7, HST-V1 is more than two magnitudes 
brighter than the turnoff; if it is a cluster member, at least
one of its components must be a giant.

\section{Summary}

The main results of our paper can be summarized as follows:

(1) The core of M5 contains a highly centrally concentrated population of 
blue stragglers, similar in size to that found in 47 Tuc (Paresce \et 1991). 

(2) Of 24 blue stragglers detected in the HST images, only one shows 
significant variability above the noise level. Since our data are not 
sensitive enough to detect variables of amplitude $\leq 0.2$ mag in the 
V=17-18 range, we cannot exclude the possibility that some blue stragglers 
may be variable below that level. A more sensitive search with the refurbished 
HST is highly desirable.

(3) We confirm the existence of a 0.2 mag gap in the color-magnitude
distribution of the horizontal branch, first noted by Brocato \et . The nature
of this gap is still unknown.

(4) We have presented finding charts and light curves for 27 RR Lyrae variables
and one probable contact binary. Among those variables, seven were previously
unknown and 21 had no published light curve.

\acknowledgments

Support for this work was provided by NASA through grant number
GO-3872 from the Space Telescope Science Institute, which
is operated by the Association of Universities for Research in Astronomy,
Inc., under NASA contract NAS5-26555.

\begin{figure}
\figurenum{TABLE 1}
\plotone{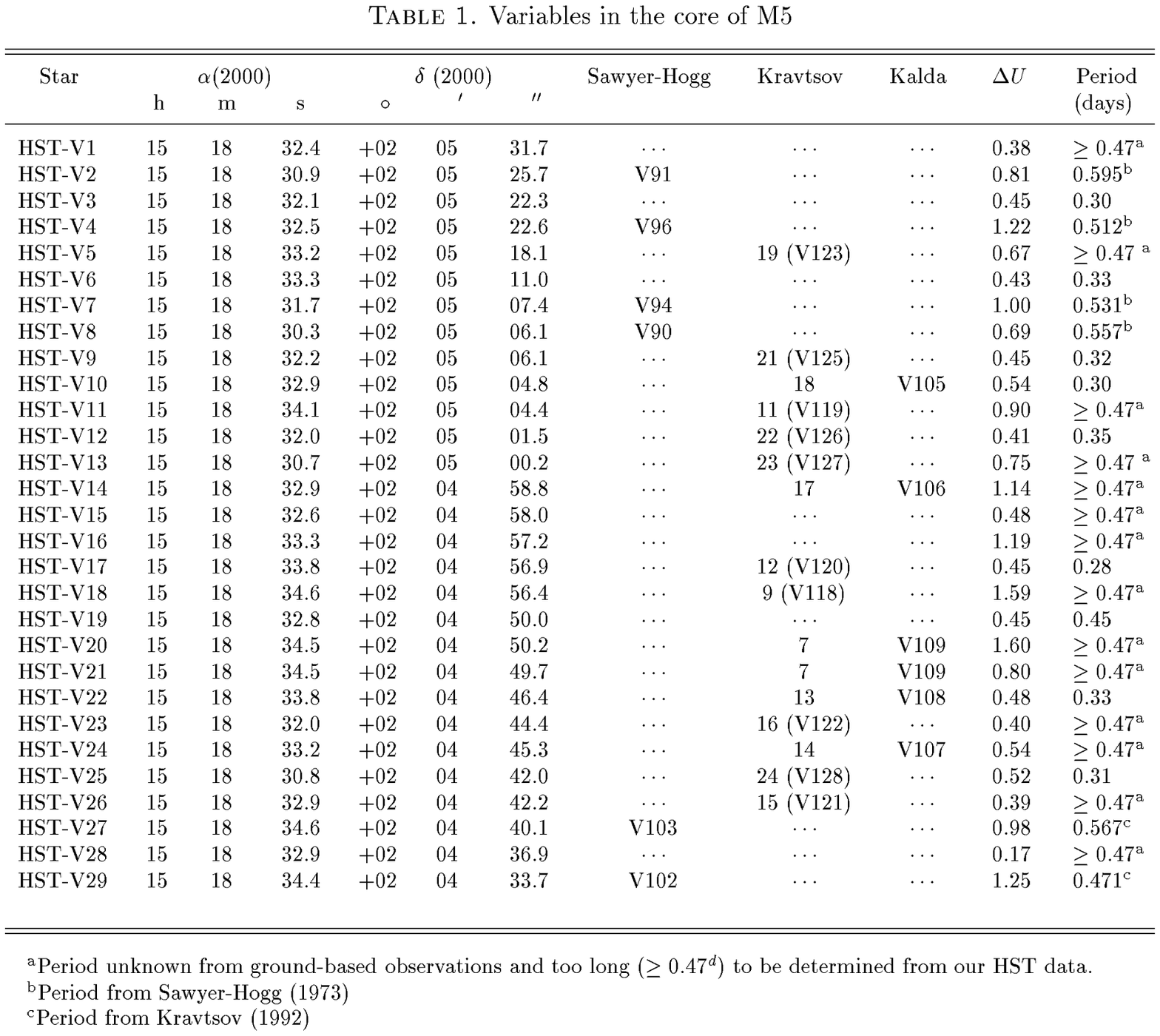}
\end{figure}

\begin{figure}
\figurenum{1}
\plotone{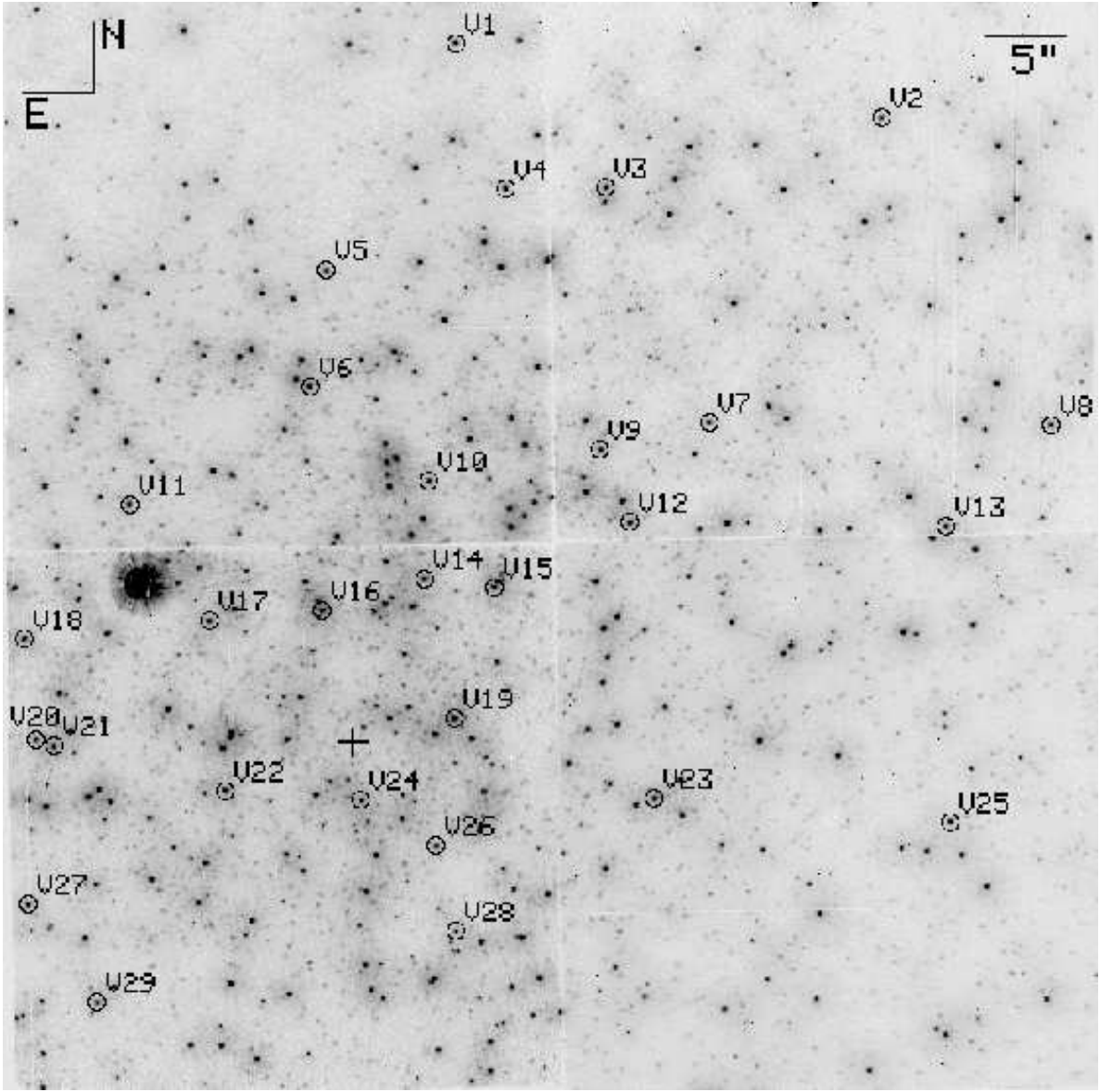}
\figcaption{ Full HST/PC field of view ($70 '' \times 70''$)
of the central region of M5. This image is the average of twenty-two
F336W (U-band) exposures. Variable stars and the cluster center
are identified.}
\end{figure}

\begin{figure}
\figurenum{2}
\plotone{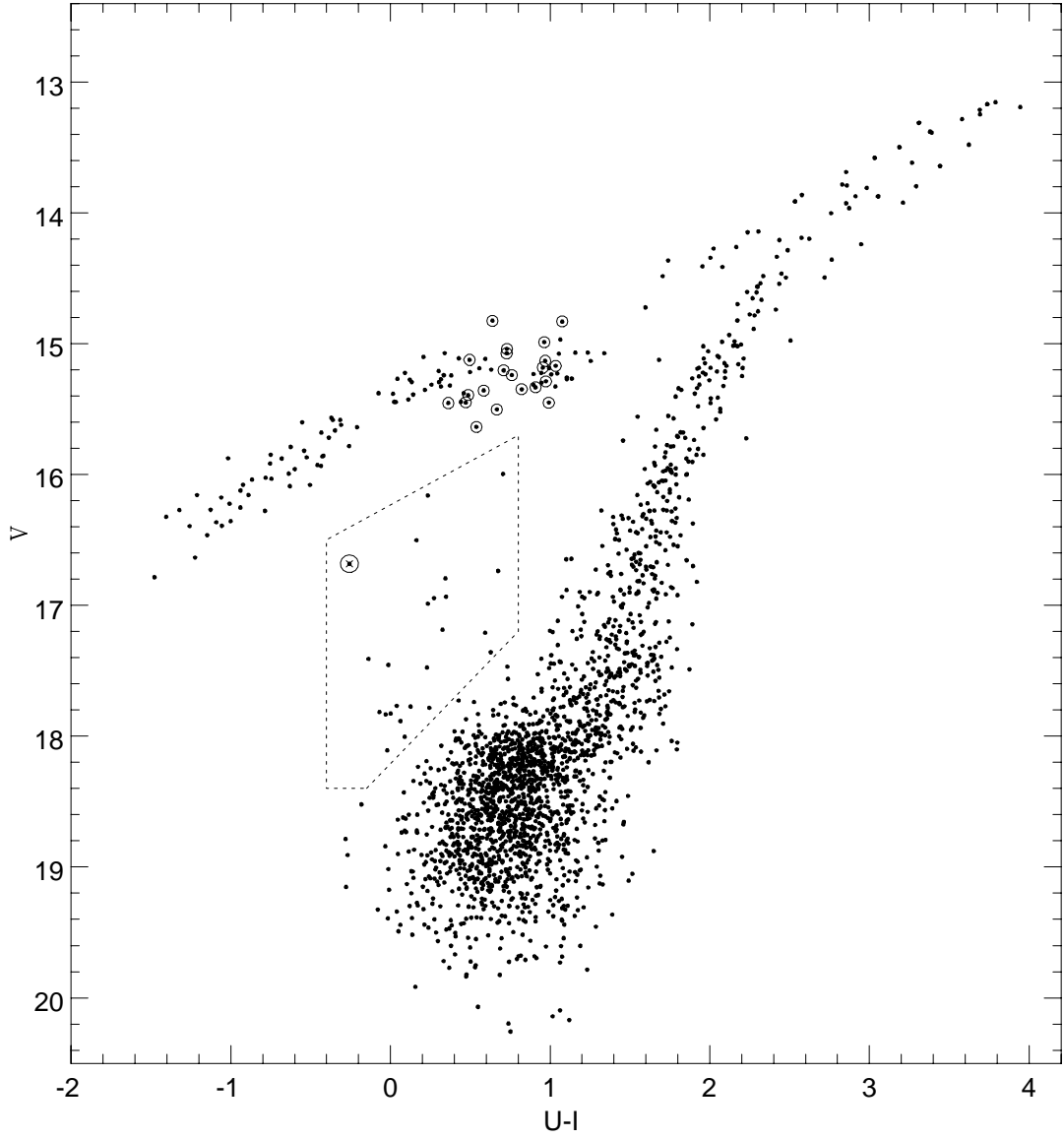}
\figcaption{Color-magnitude diagram for the stars common to the
U (F336W), V(F555W) and I(F785LP) frames. RR Lyrae stars, and the
variable blue straggler are shown with special symbols.}
\end{figure}

\begin{figure}
\figurenum{3}
\plotone{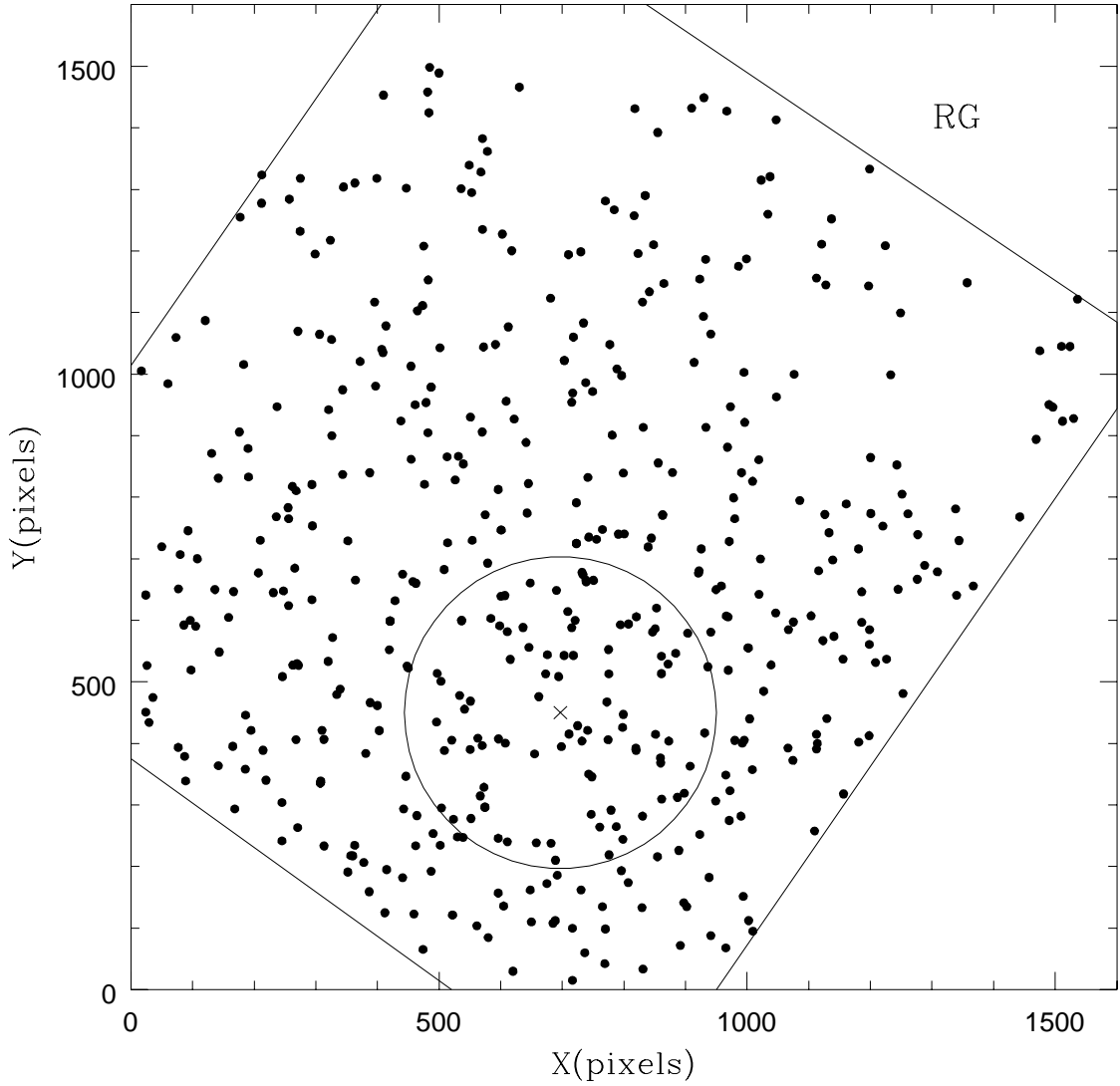}
\figcaption{Distribution of the red giants, horizontal branch stars
and blue stragglers in the V frame coordinate system. The location of
the U images is shown. The cluster center is identified by a cross.
The circle has a radius of $20''$ ($\sim$ one core radius).}
\end{figure}

\begin{figure}
\figurenum{3}
\plotone{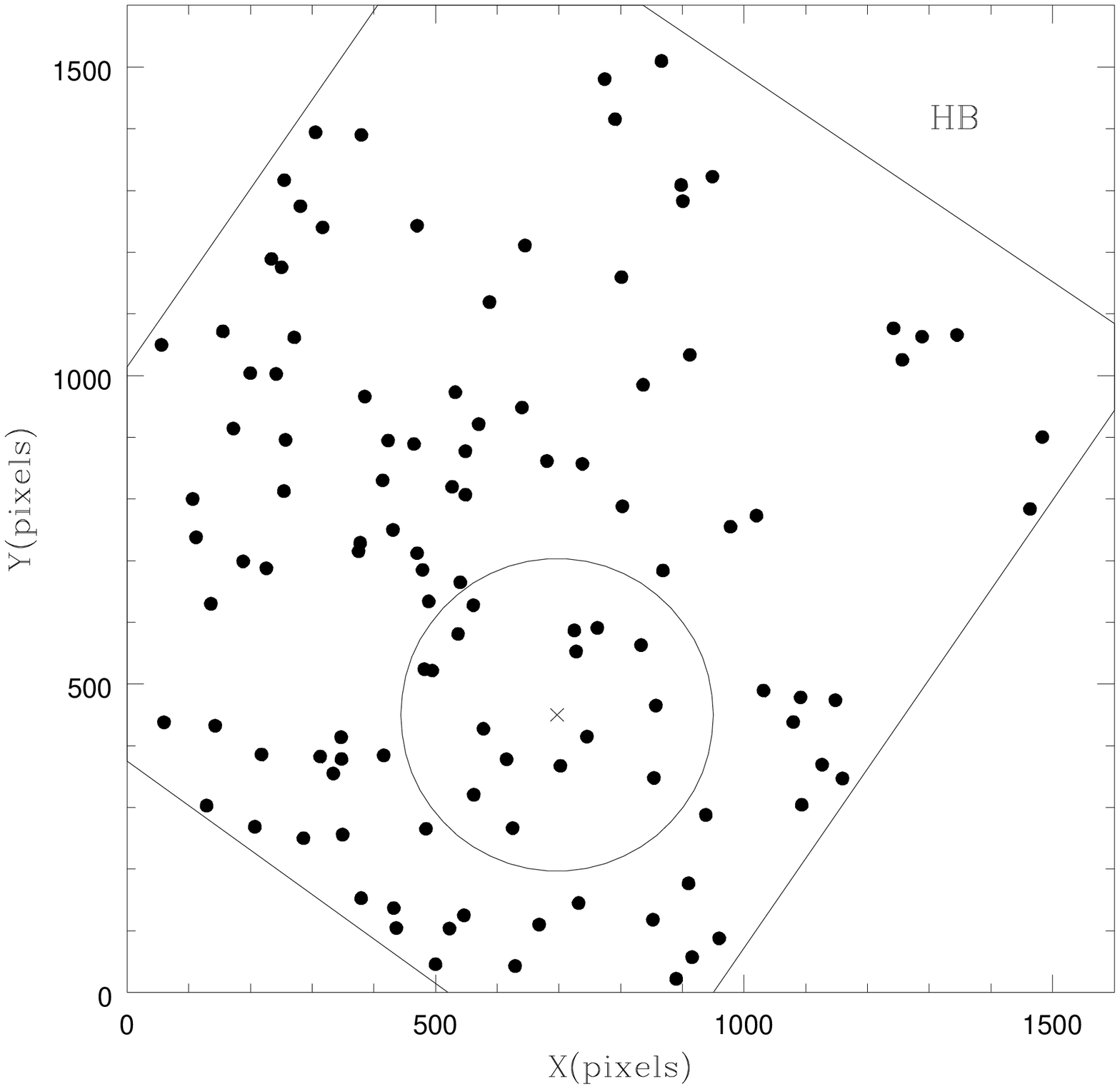}
\figcaption{continued}
\end{figure}

\begin{figure}
\figurenum{3}
\plotone{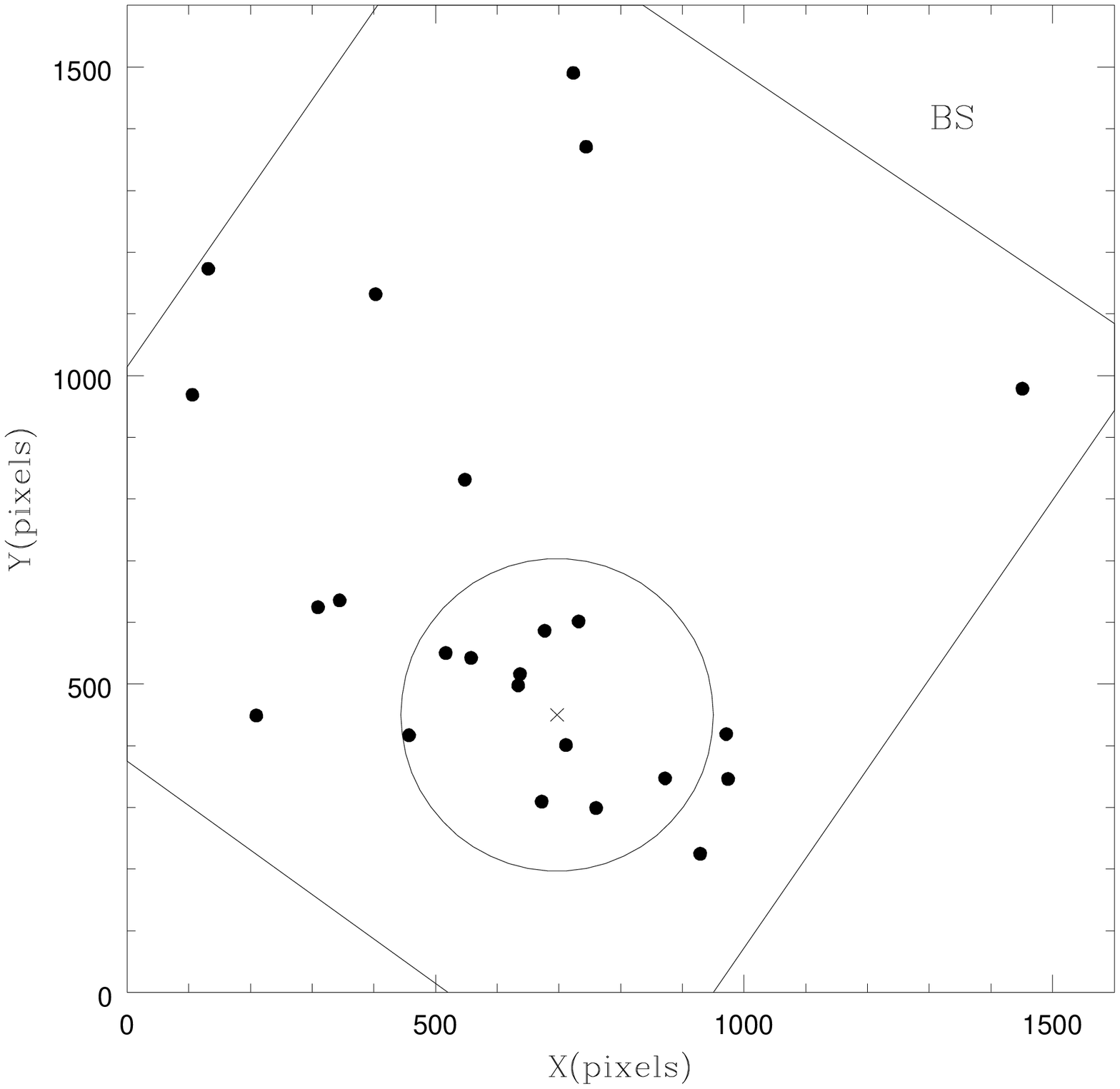}
\figcaption{continued}
\end{figure}

\begin{figure}
\figurenum{4a}
\plottwo{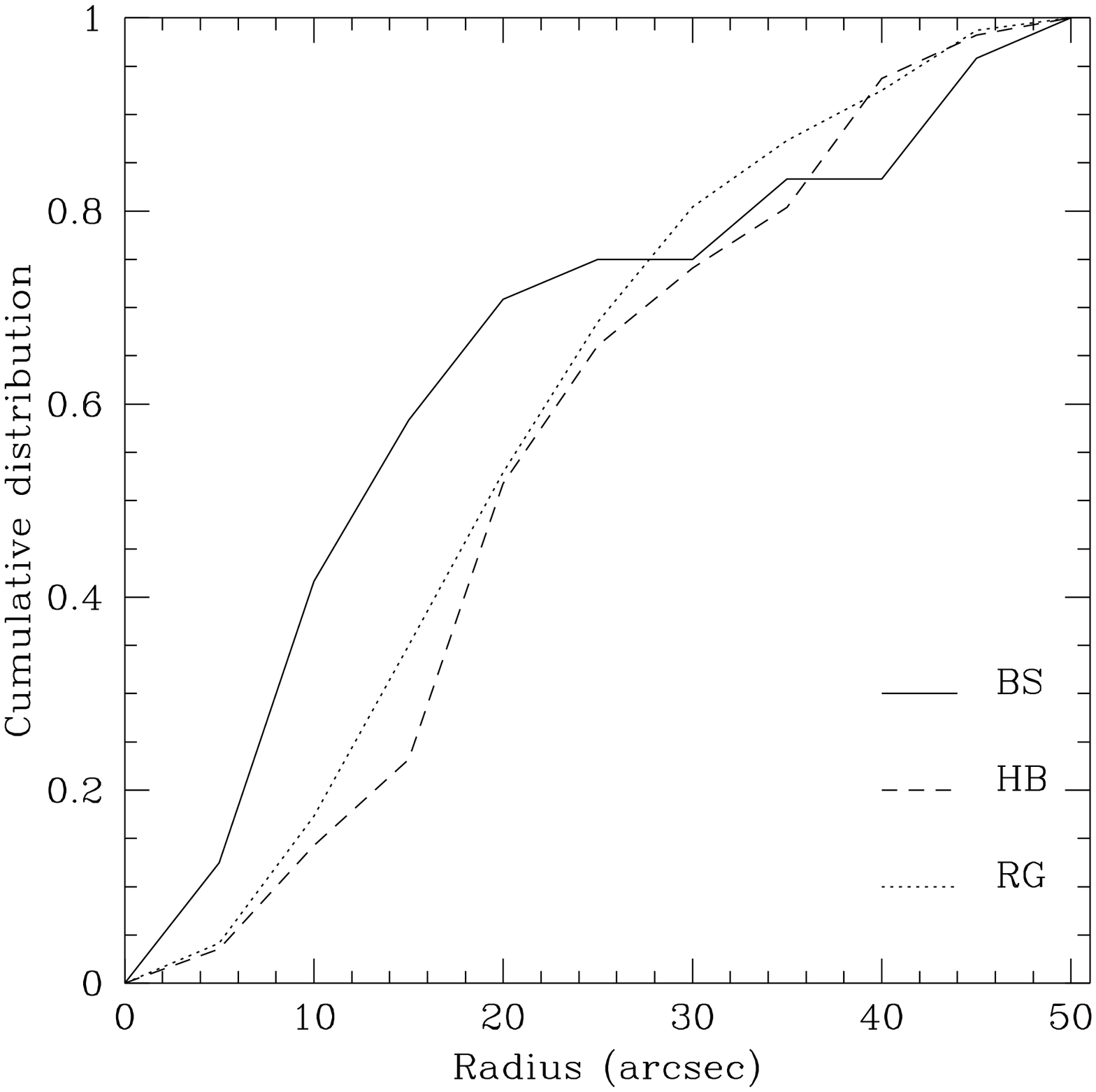}{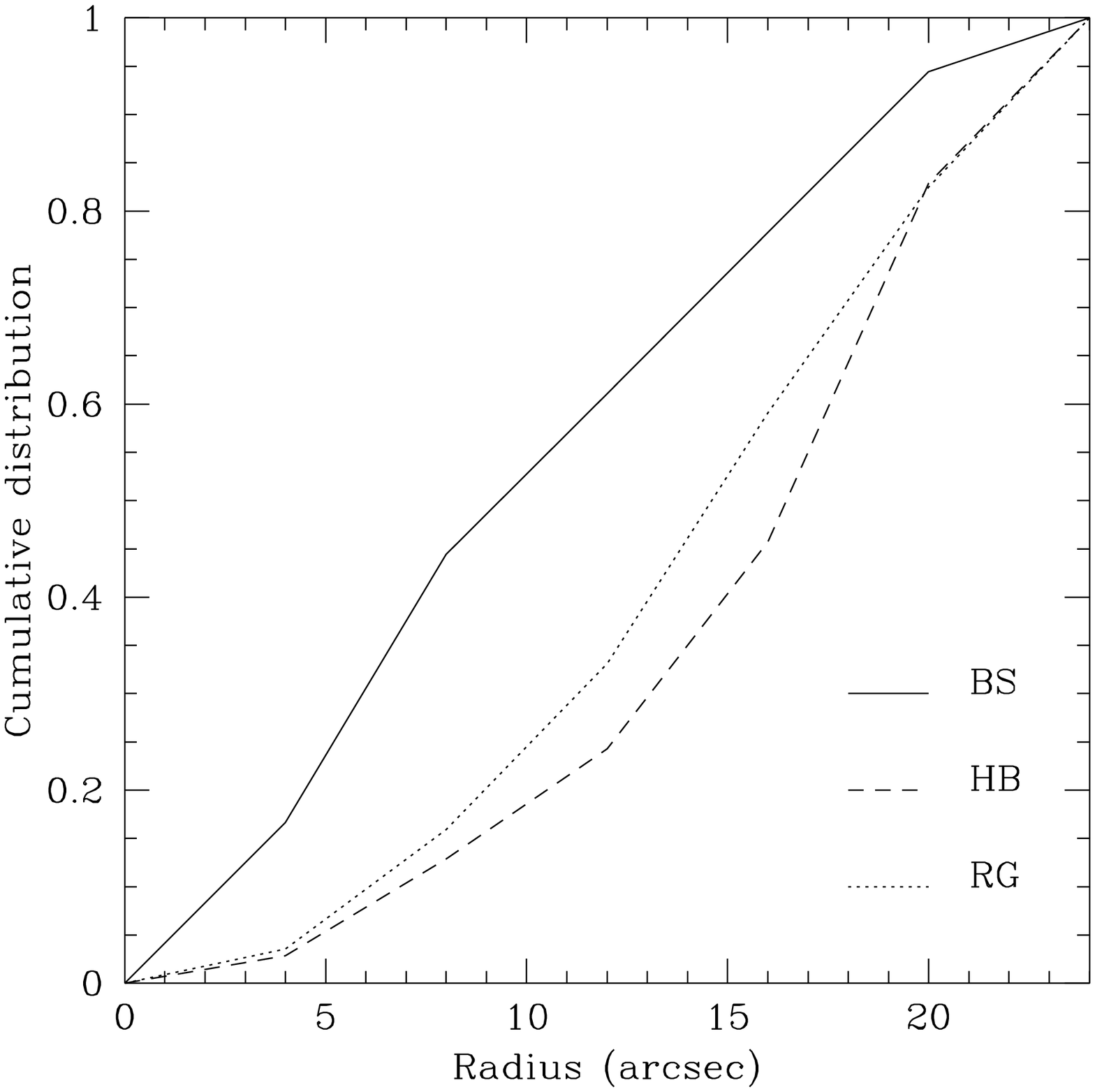}
\figcaption{(a) Cumulative distribution of the stellar populations
(BS: Blue Stragglers; HB: Horizontal Branch stars; RG: Red Giants) in
the inner 50$''$ region of M5; (b) same as (a), but within one core radius}
\end{figure}

\begin{figure}
\figurenum{5}
\plotone{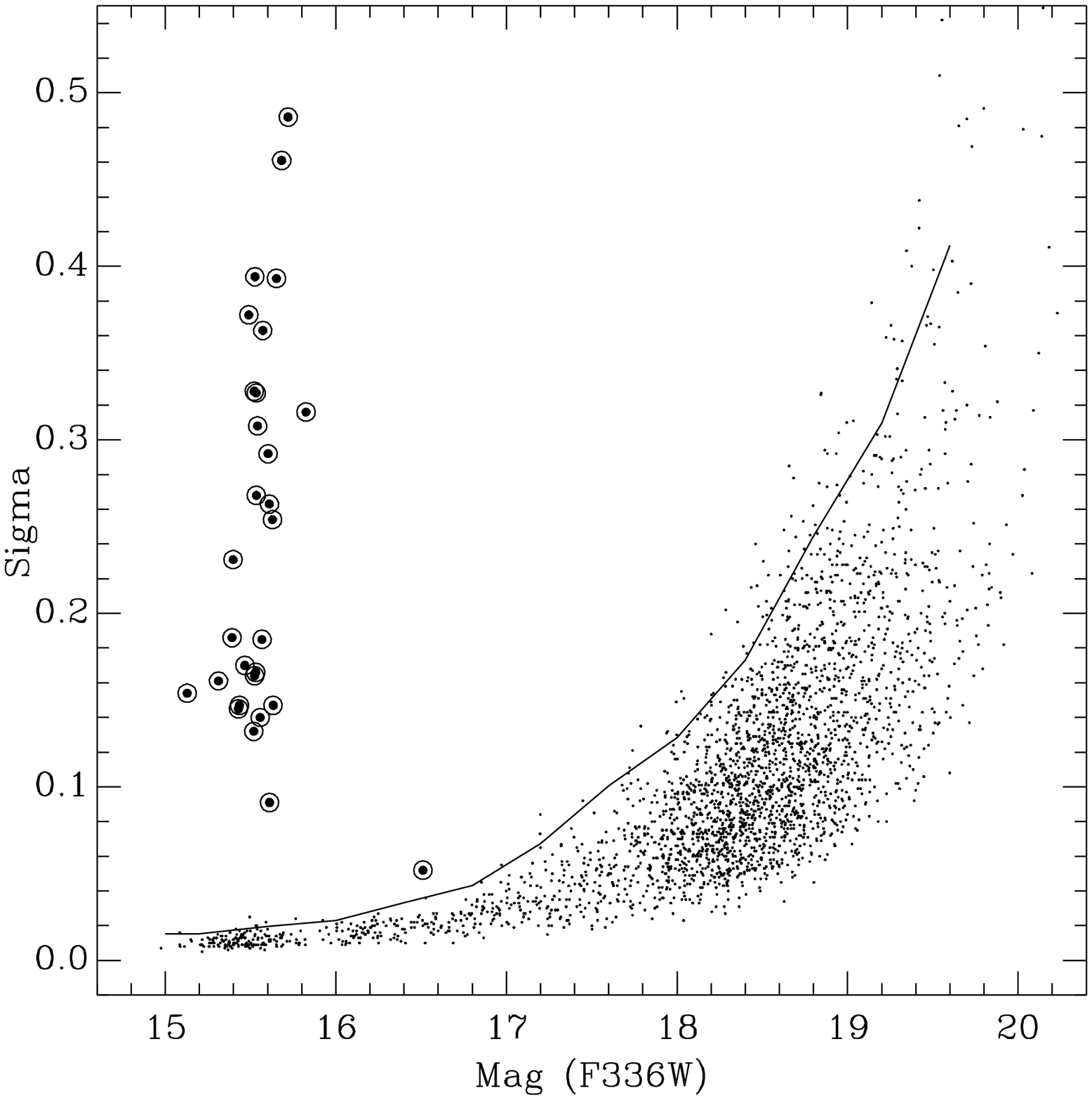}
\figcaption{The RMS magnitude of variability as a function of the
F336W magnitudes for the stars in the 21 frame set. The line is the 2$\sigma$
cutoff, above which the stars were considered as candidate variables and
examined more closely. Genuine variables are shown with special symbols.}
\end{figure}

\begin{figure}
\figurenum{6}
\plotone{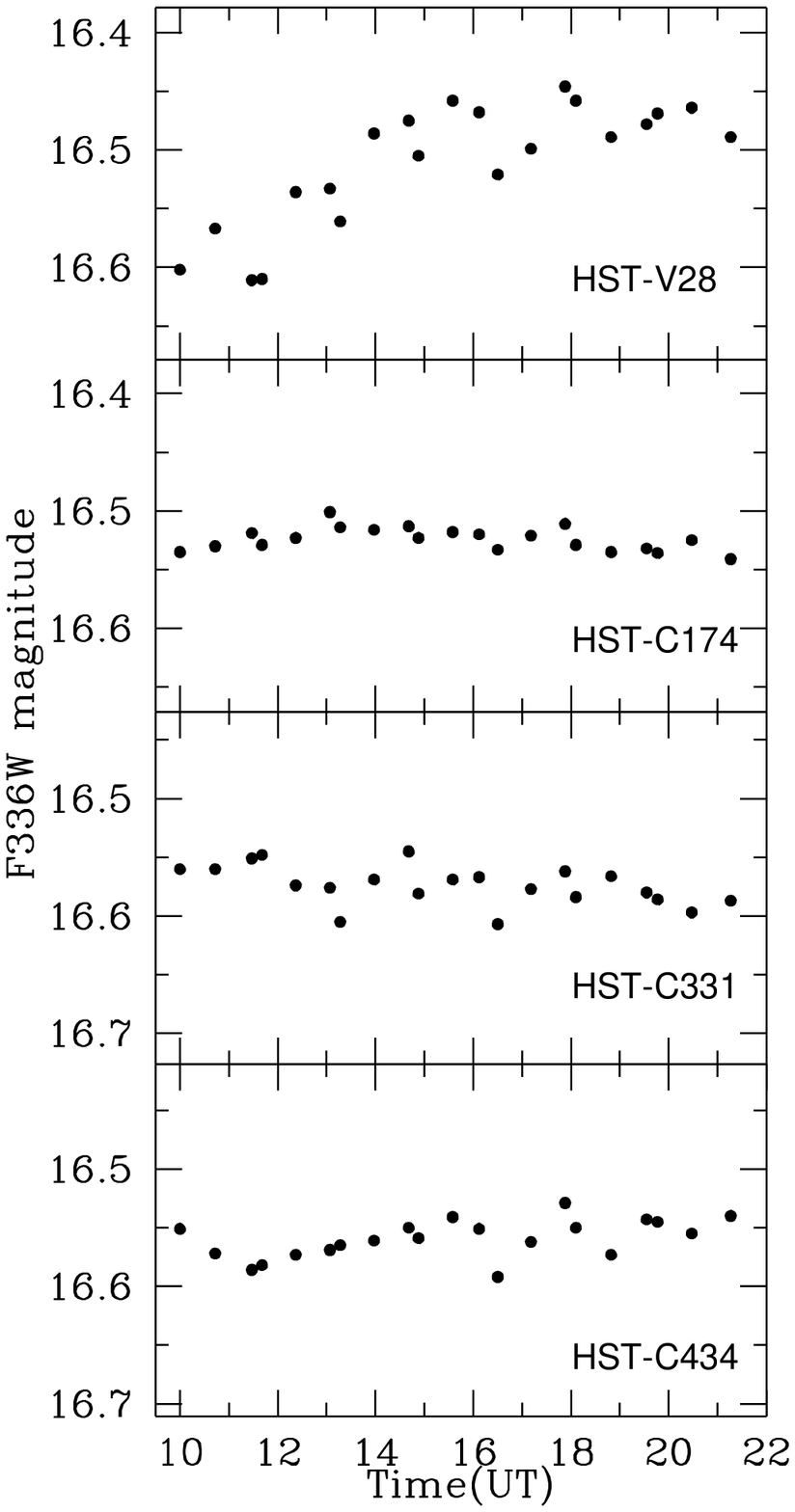}
\figcaption{Light curve of HST-V28, the variable blue straggler
and of three nearby ($\Delta d \leq 5''$)
comparison stars of similar mean magnitude}
\end{figure}

\begin{figure}
\figurenum{7}
\plotone{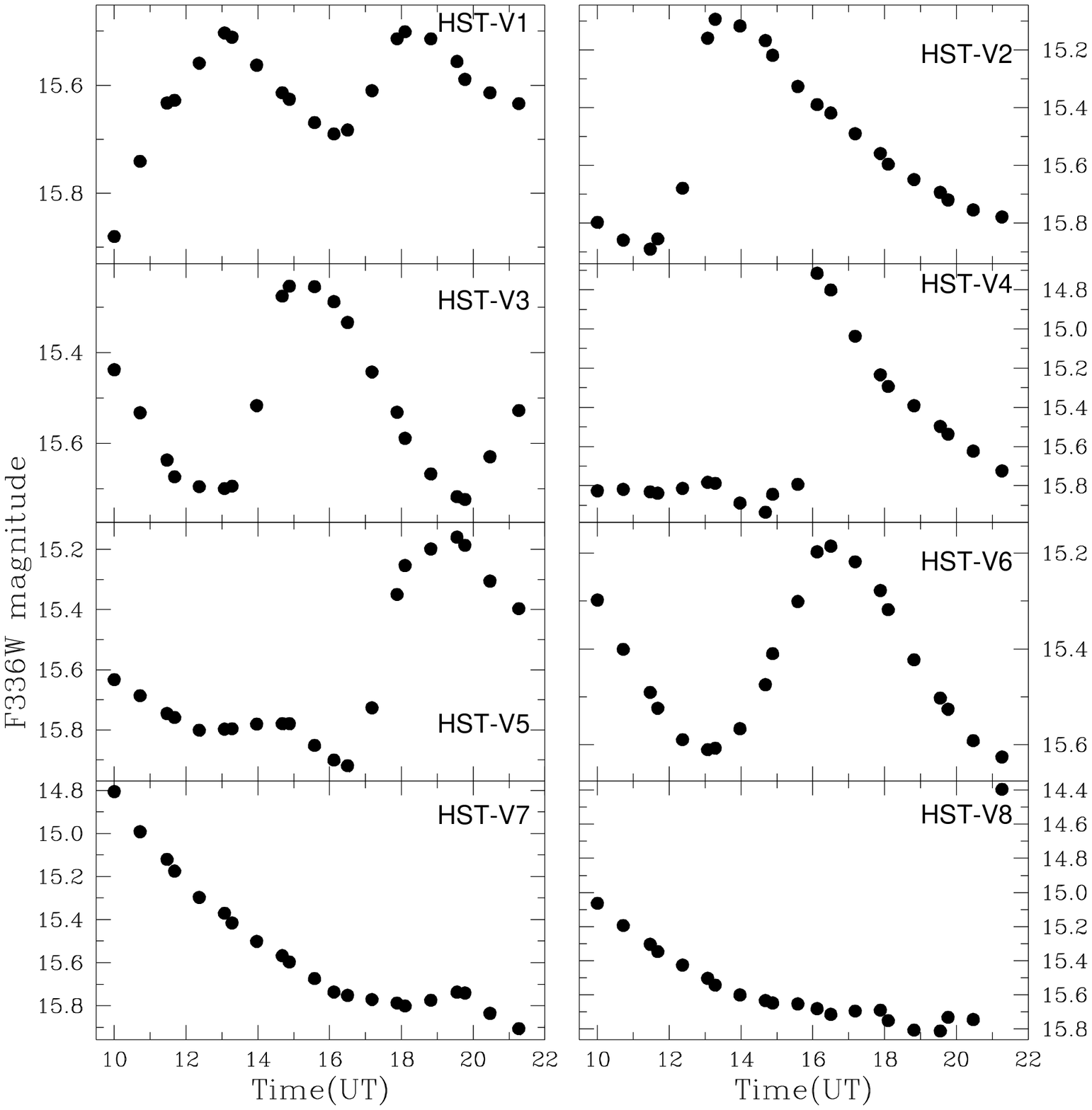}
\figcaption{Light curves (filter F336W $\sim$ U-band) of the large amplitude 
variable stars found in the core of M5. All but one (HST-V1) are RR Lyrae 
stars.}
\end{figure}

\begin{figure}
\figurenum{7}
\plotone{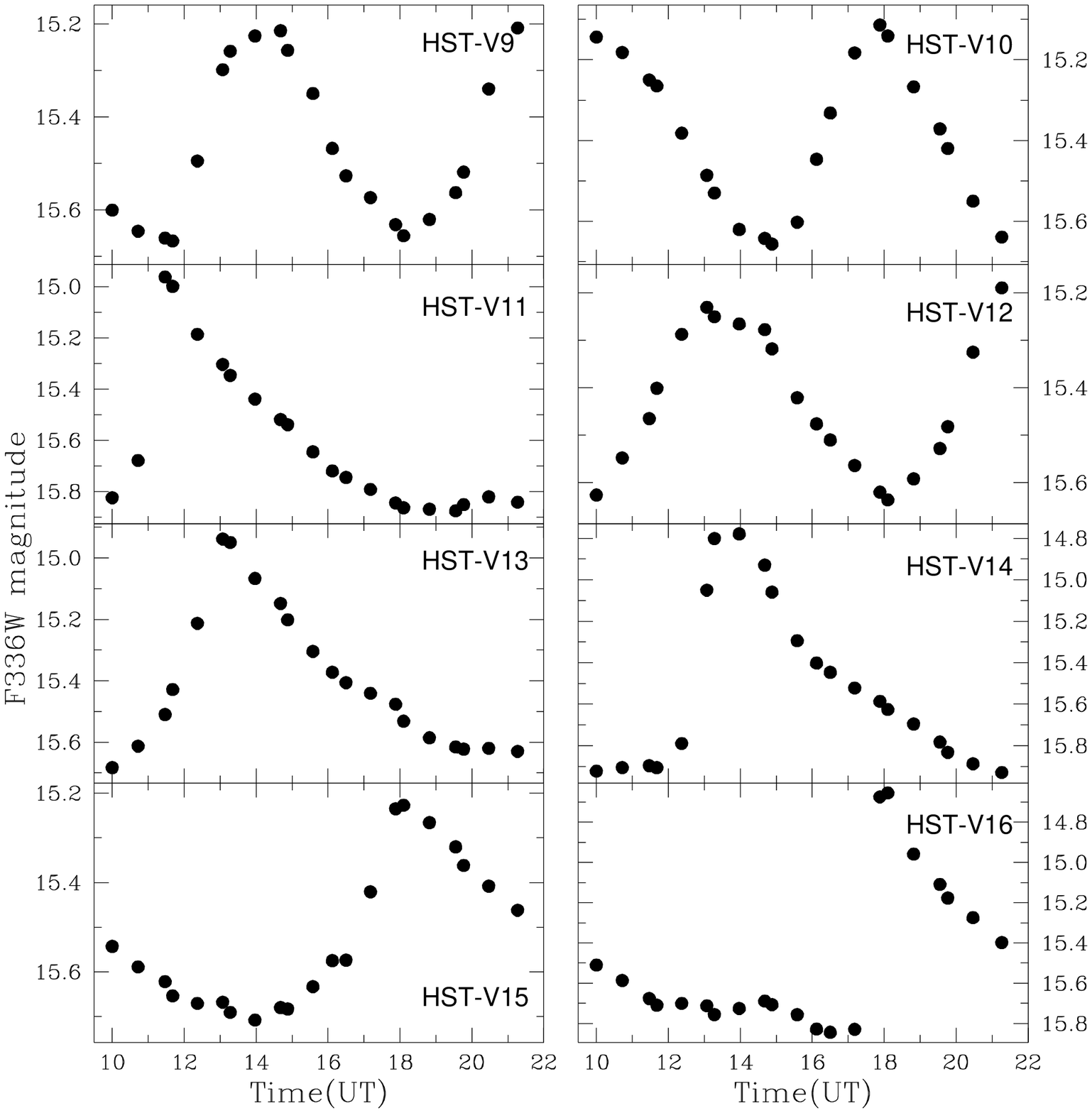}
\figcaption{Continued}
\end{figure}

\begin{figure}
\figurenum{7}
\plotone{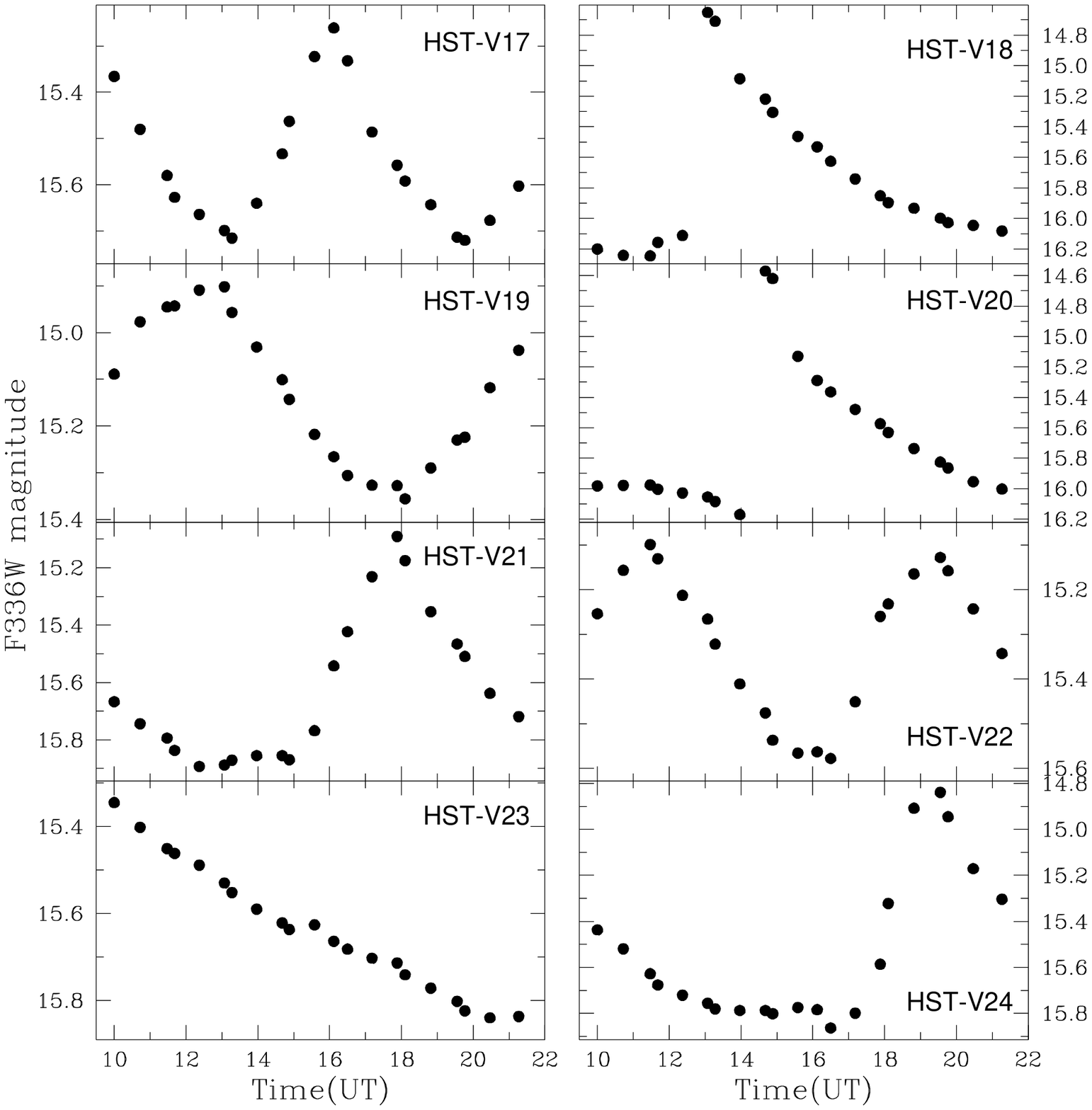}
\figcaption{Continued}
\end{figure}

\begin{figure}
\figurenum{7}
\plotone{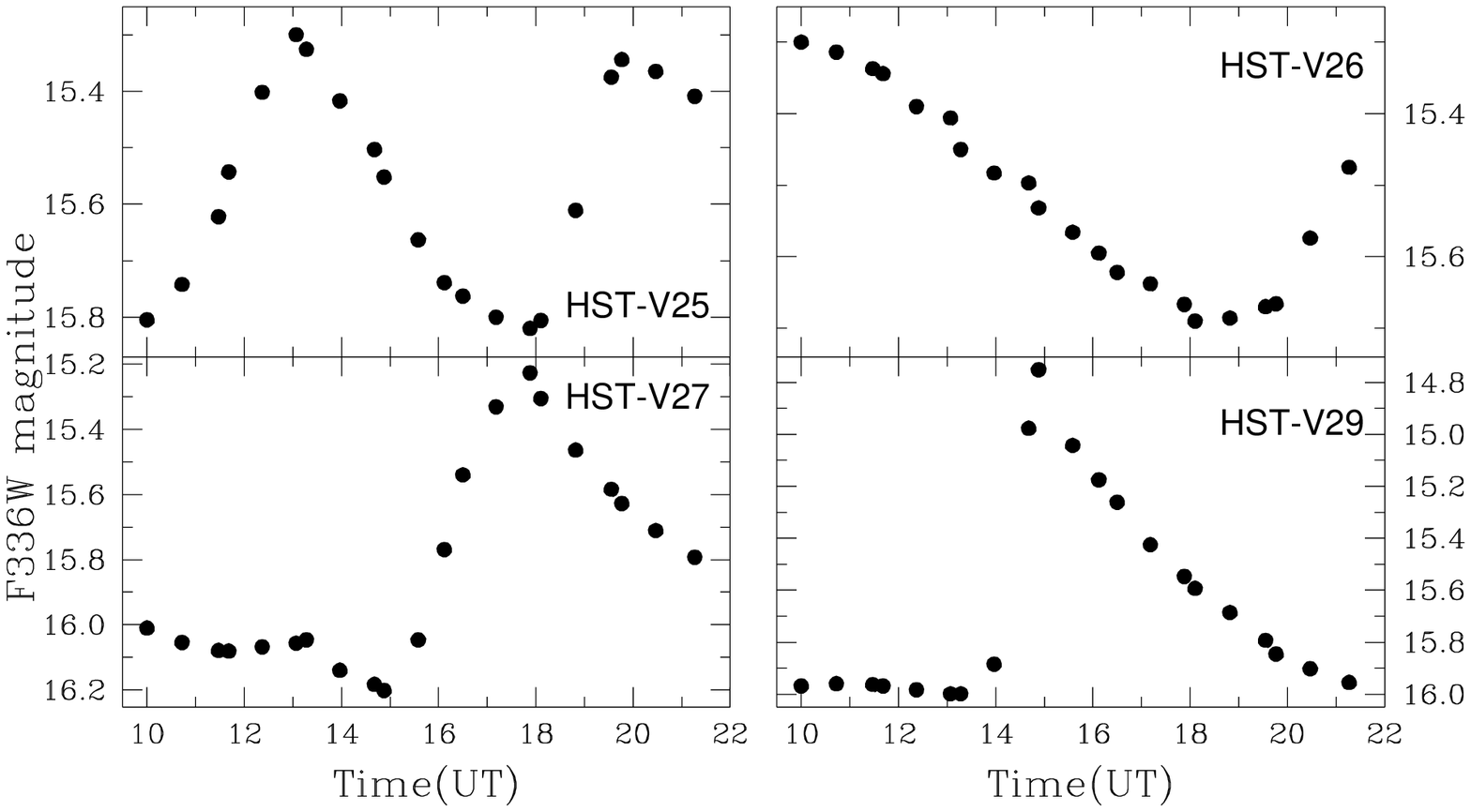}
\figcaption{Continued}
\end{figure}

\end{document}